%% file: main.tex
\documentclass[%
 pra,aps,
 amsmath,amssymb,
 reprint,nofootinbib,
 superscriptaddress,
 floatfix,%
]{revtex4-2}

\usepackage[utf8]{inputenc}
\usepackage{graphicx}
\usepackage{physics}
\usepackage[caption=false]{subfig}
\usepackage{xcolor}
\usepackage{pgfplots}
\usepackage{tikz}
\usetikzlibrary{patterns}
\pgfplotsset{compat=1.11,
    /pgfplots/ybar legend/.style={
    /pgfplots/legend image code/.code={%
       \draw[##1,/tikz/.cd,yshift=-0.25em]
        (0cm,0cm) rectangle (3pt,0.8em);},
   },
}
\allowdisplaybreaks
\usepackage[acronym, nomain]{glossaries} 
\usepackage{tabularx}
\usepackage[nodisplayskipstretch]{setspace}

\usepackage[bookmarks=false]{hyperref}
    \hypersetup{colorlinks,
      linkcolor=blue,
      citecolor=blue,
      urlcolor=blue}

\input{acronyms}

\begin{document}

\author{Moritz Willmann}
\email{moritz.willmann@ipa.fraunhofer.de}
\affiliation{Fraunhofer Institute for Manufacturing Engineering and Automation IPA, Nobelstraße 12, D-70569 Stuttgart, Germany}
\affiliation{Institute of Industrial Manufacturing and Management IFF, University of Stuttgart, Allmandring 35, D-70569 Stuttgart, Germany}

\author{Marcel Albus}
\email{marcel.albus@ipa.fraunhofer.de}
\affiliation{Fraunhofer Institute for Manufacturing Engineering and Automation IPA, Nobelstraße 12, D-70569 Stuttgart, Germany}
\affiliation{Institute of Industrial Manufacturing and Management IFF, University of Stuttgart, Allmandring 35, D-70569 Stuttgart, Germany}

\author{Jan Schnabel}
\author{Marco Roth}
\affiliation{Fraunhofer Institute for Manufacturing Engineering and Automation IPA, Nobelstraße 12, D-70569 Stuttgart, Germany}

\title{Application of quantum annealing for scalable robotic assembly line optimization: a case study} 

\date{October 23, 2024}

\begin{abstract}
	The even distribution and optimization of tasks across resources and workstations is a critical process in manufacturing aimed at maximizing efficiency, productivity, and profitability, known as \gls{ralb}.
	With the increasing complexity of manufacturing required by mass customization, traditional computational approaches struggle to solve \gls{ralb} problems efficiently.
	To address these scalability challenges, we investigate applying quantum computing, particularly quantum annealing, to the real-world based problem.
	We transform the integer programming formulation into a quadratic unconstrained binary optimization problem, which is then solved using a hybrid quantum-classical algorithm on the \textit{D-Wave Advantage 4.1} quantum computer.
	In a case study, the quantum solution is compared to an exact solution, demonstrating the potential for quantum computing to enhance manufacturing productivity and reduce costs.
	Nevertheless, limitations of quantum annealing, including hardware constraints and problem-specific challenges, suggest that continued advancements in quantum technology will be necessary to improve its applicability to \gls{ralb} manufacturing optimization.
\end{abstract}


\maketitle

\section{Introduction}
\label{sec:introduction}
\glsresetall 

\gls{ralb} is crucial for maximizing production efficiency, productivity, and profitability~\cite{Becker2006}.
It involves assigning tasks among workstations to optimize production time or workstation count, considering automated resources and alternatives~\cite{Rubinovitz1993}.
An optimal line balance is essential for today's manufacturing companies to meet customer demands while maintaining competitiveness~\cite{Touzout2019}.
With increasing manufacturing complexity due to mass customization, traditional approaches face scalability challanges and struggle to solve the \glspl{ralbp} efficiently~\cite{Riandari2021, Boysen2022}.

Quantum computing shows promising results in addressing these scalability challenges, potentially enabling near-optimal or faster optimal solutions for \gls{ralbp}~\cite{Luckow2021, Pirnay2024}.
Quantum algorithms like \gls{qaoa}~\cite{Farhi2014}, and quantum annealing~\cite{Finnila1994} have shown promising results in solving different optimization problems~\cite{Abbas2023, King2023}, making them appealing for the \gls{ralbp}.

We investigate the potential of quantum computing for \gls{ralbp}, scrutinizing the approach to address scalability challenges by translating the \gls{ip} problem into a \gls{qubo} and mapping it to an Ising model.
We solve a problem instance on the \textit{D-Wave Advantage 4.1} quantum computer to identify potentials and problems.
In this use case study, we compare a quantum approach to classical methods for solving the \gls{ralbp}, assessing their practical benefits in real-world manufacturing scenarios. 
By detailing the procedure to address the problem with quantum computing and providing a corresponding python tool~\cite{alb_qubo_code}, we aim at enabling manufacturers to assess the potential of quantum computing for the \gls{ralbp}.

The remainder of this paper is structured as follows.
In \autoref{sec:related_work}, the current state of the art of solving the \gls{ralbp} as well as the application of quantum computing to \glspl{cop} is discussed.
\autoref{sec:methods} presents the \gls{ip} and the \gls{qubo} model formulation, and the quantum annealing solution approach.
A \gls{ralbp} case study is presented in \autoref{sec:case_study}, and \autoref{sec:discussion} discusses the results and provides an outlook.

\section{Related Work}
\label{sec:related_work}

The definition of the \gls{ralbp}, is driven by real-world manufacturing needs for flexibility and increased competitiveness through specialized production equipment~\cite{Rubinovitz1993}.
Classical methods for solving the \gls{ralbp} include exact methods~\cite{Li2020}, heuristics~\cite{Zhang2018, Borba2018} and meta-heuristic approaches~\cite{Albus2024, Manavizadeh2012}, with significant improvements in the last decades~\cite{Chutima2022}.
However, these methods often struggle when dealing with large-scale and highly complex instances~\cite{Klar2022}.
Either they provide high-quality results for small instances through exact solution techniques~\cite{Dolgui2009}, or scalability in complex instances with reduced quality through meta-heuristic techniques~\cite{Albus2024}.
As the number of tasks and workstations increases, the computational complexity proves a significant issue due to the NP-hard problem type, making it infeasible to find optimal solutions within a reasonable time.
This scalability issue has motivated the exploration of new computational paradigms, such as quantum computing, to tackle the \gls{ralbp} more efficiently.

Recently, the availability of early-stage quantum computing devices has raised interest in applying quantum algorithms to a wide range of problems.
Alongside the simulation of physical systems and machine learning, \glspl{cop} has been one of the areas of research where quantum computing has been applied \cite{Abbas2023}.
Quantum annealing, a specialized form of quantum computation, is naturally suited for solving \glspl{cop}~\cite{Yarkoni2022}.
It returns the lowest possible energy in a physical system described by the Ising Hamiltonian, which corresponds to the optimal solution of the \gls{cop}.
Quantum annealers are already commercially available and have been applied to various industrial problems~\cite{Cohen2014, Liu2021}.

Quantum annealing has shown promising results in manufacturing, including layout planning~\cite{Klar2022}, job shop scheduling~\cite{Kurowski2020}, and production planning~\cite{Riandari2021}.
A list of industry-related reference problems suitable for quantum computing in the automotive industry has been curated~\cite{Schworm2023}, with some already being addressed~\cite{Glos2023}.
Benefits of quantum annealing in manufacturing include handling complex combinatorial problems, potentially finding higher-quality solutions, or faster solution generation for certain problems~\cite{Schworm2023,Yarkoni2022,Luckow2021}.
However, limited qubit counts, connectivity issues, and noise in quantum systems currently restrict its universal applicability.

\section{Methods}
\label{sec:methods}

In this section, we introduce the \gls{ip} model for the \gls{ralbp}.
This is then used to reformulate the problem in a \gls{qubo} form which allows for a solution with a quantum computer.

\subsection{Integer Programming Model Formulation}
\label{subsec:ip_model_formulation}

The \gls{ralbp} aims to assign tasks to production equipment and allocate equipment to workstations to manufacture a desired product.
An assignment optimization can maximize efficiency or minimize investment costs.
Manufacturing requires all tasks to be assigned, each with equipment-specific processing times, following a predetermined assembly order within a given cycle time.
The pieces of equipment are selected from available production resources.
The \gls{ip} model formulation requires two binary decision variables, one for the assignment of equipment to workstations and the second for the assignment of tasks to workstations.
We define $y_{jk}=1$ if equipment $j$ is assigned to workstation $k$, $0$ otherwise.
Additionally, $x_{ijk}=1$ if task $i$ is performed by equipment $j$ at workstation $k$, $0$ otherwise.
An overview of the variable notation for the \gls{ralbp} model is given in \autoref{tab:variable_description}.

This results in the \gls{ip} problem formulation for the \gls{ralbp}~\cite{Rubinovitz1993}:
\begin{align}
	& \min \sum_{j=1}^{r}\sum_{k=1}^{m} c_{j} y_{jk}                                              &   & \label{eq:objective}                                              \\
	& \text{s.t.}                                                                                 &   & \nonumber                                                         \\
	& \sum_{j=1}^{r} \sum_{k=1}^{m} x_{ijk} = 1                                               &   & \forall i \label{eq:constraint:task}                              \\
	& \sum_{i=1}^{n} t_{ij} x_{ijk} \leq C y_{jk}                                                 &   & \forall j, k \label{eq:constraint:task_time}                      \\
	& \sum_{i=1}^{n} \sum_{j=1}^{r} t_{ij} x_{ijk} \leq C                                         &   & \forall k \label{eq:constraint:workstation_time}                  \\
	& \sum_{j=1}^{r} \sum_{k=1}^{m} k x_{pjk} \leq \sum_{j=1}^{r}\sum_{l=1}^{m} l x_{ijl} &   & \forall p \in \mathcal{P}_{i}, i  \label{eq:constraint:precedence}
\end{align}
The objective function \eqref{eq:objective} minimizes total equipment costs.
Constraint \eqref{eq:constraint:task} assigns each task to exactly one workstation.
Constraint \eqref{eq:constraint:task_time} ensures that the task time at each piece of equipment does not exceed cycle time and that pieces of equipment can only perform tasks if assigned to the workstation.
Additionally, multiple pieces of equipment can be assigned to one workstation.
The sum of the total workstation task time is limited to cycle time by constraint \eqref{eq:constraint:workstation_time}.
Constraint \eqref{eq:constraint:precedence} enforces task precedence graph relations given by edges \mbox{$\mathcal{E} = \lbrace (p, i)\,\vert\, p \in \mathcal{P}_{i}, i\in[1,n]\rbrace$}, where task $p$ precedes task $i$.

\renewcommand{\arraystretch}{1.0}
\begin{table}[tb]
	\caption{Variable description and overview for the \acrlong{ip} model.}
	\label{tab:variable_description}
	\begin{tabularx}{\hsize}{@{}l X@{}}
		\hline
		Variable name     & Variable description                                                                                                                                    \\
		\hline
		$n$               & Number of tasks.                                                                                                                                        \\
		$r$               & Number of equipment.                                                                                                                                    \\
		$m$               & Number of workstations.                                                                                                                                 \\

		$i$               & Task index, $i \in \{1, \dots, n\}$.                                                                                                                    \\
		$j$               & Equipment index, $j \in \{1, \dots, r\}$.                                                                                                               \\
		$k$               & Workstation index, $k \in \{1, \dots, m\}$.                                                                                                             \\

		$t_{ij}$          & Processing time of task $i$ when performed by equipment $j$, in s.                                                                                      \\
		$C$               & Cycle time, in s.                                                                                                                                       \\
		$c_{j}$           & Cost of equipment $j$, in \$.                                                                                                                           \\



		$\mathcal{P}_{i}$ & The set of tasks which precede task $i$.                                                                                                                \\
  	\hline
	\end{tabularx}
\end{table}
\renewcommand{\arraystretch}{1.0}

\subsection{Quadratic Unconstrained Binary Optimization}
\label{subsec:qubo}

To solve the \gls{ralbp} with quantum computing, the problem needs to be formulated as a \gls{qubo}.
\gls{qubo} is a problem class that is capable of representing a wide range of \glspl{cop} as minimization problems of the general form
\begin{equation}
	\label{eq:qubo}
	\min_{x \in \{0,1\}^n} x^\intercal Qx \,,
\end{equation}
with the $n$-dimensional decision vector $x$ and the quadratic \gls{qubo}-Matrix $Q$~\cite{Lucas2014}.
In the case of the \gls{ralbp}, the optimization problem is formulated as a linear integer program with constraints, which gets transformed with the following steps.

First, the linear cost function of the \gls{ralbp} in Eq.~\eqref{eq:objective} can be easily transformed to a quadratic one for the binary case because the equality $x = x^2$ holds for $x \in \{0, 1\}$.
Second, equality constraints of the form $Ax=b$ can be reformulated and added to the quadratic cost function as
\begin{equation}
	\label{eq:reformulation:lagrange}
	\min_{x \in \{0,1\}, Ax=b} x^\intercal Q x=\min_{x \in \{0,1\}} x^\intercal Qx+\lambda\left( Ax-b \right)^2\,,
\end{equation}
with help of Lagrange-parameters $\lambda$.
In the case of the \gls{ralbp}, four Lagrange parameters are added, one for each constraint.
This transformation changes the hard constraints of the original problem formulation to soft constraints.
Third, inequality constraints can be transformed into equality constraints, as in \mbox{$Ax \leq b \Leftrightarrow Ax + s = b, s>0$}, by introducing slack variables $s$.
Technically, the slack variables are implemented as decision variables with weights that can produce values between $0$ and $b$, \ie, the binary representation of $b$.
Their use increases the problem size significantly.
The amount of additional decision variables are determined by the binary representation.
In total, these transformations introduce \mbox{$n_s = (r + rm) \lceil \log_2(C) \rceil + \abs{\mathcal{E}} \lceil \log_2(nrm) \rceil$} new variables.
The first term originates from the two inequality constraints Eq.~\eqref{eq:constraint:task_time} and Eq.~\eqref{eq:constraint:workstation_time} that include the cycle time $C$ in the original formulation.
The second term stems from the inequality constraint Eq.~\eqref{eq:constraint:precedence} of the precedence graph, with $\abs{\mathcal{E}}$ the number of edges.
To reduce the total number of decision variables, rescaling the cycle time $C$ and all the task times $t_i$ by dividing by their greatest common divisor is advisable.
Additionally, one only needs to consider the tasks assigned to machines permitted by $t_{ij}$ to reduce the number of decision variables needed.
The transformation process also requires fixing the number of workstations $m$ to yield a constant \gls{qubo} size.

To automatically transform the \gls{ralbp} into a \gls{qubo} representation, we have developped a python library which is available in GitLab \cite{alb_qubo_code}.
The library allows the formulation of arbitrary \gls{ralbp} instances and their corresponding \gls{qubo} representations, which then can be solved with any available \gls{qubo} solver.

\subsection{Quantum Annealing}
\label{subsec:quantum_annealing}

The \gls{qubo} problem can be mapped onto a physical model describing interacting qubits called Ising model.
Finding the ground state of this model is then equivalent to solving the \gls{qubo}.
The Ising model is given by the Hamiltonian
\begin{equation}
	\label{eq:ising}
 	H_\text{Ising} = -\sum_{i<j} J_{ij} \hat{\sigma}^z_i \hat{\sigma}^z_j - \sum_j h_j \hat{\sigma}^z_j \,.
\end{equation}
Here, $\hat{\sigma}^z_j$ is the Pauli $Z$-operator that describes the state of qubit $j$ in the computational basis.
The terms $J_{ij}$ and $h_j$ correspond to the interaction strength between qubits and an external field.
This Hamiltonian can be directly implemented on a quantum computer, making the problem susceptible to a variety of algorithms designed to find a system's minimum energy configuration.

To map Eq.~\eqref{eq:qubo} onto Eq.~\eqref{eq:ising}, we apply the transformation $\hat{\sigma}^z_j = 1 - 2 x_j$.
The diagonal elements $Q_{jj}$ of the \gls{qubo}-matrix in Eq.~\eqref{eq:qubo} are identified with the linear terms $h_j$ in Eq.~\eqref{eq:ising} while the off-diagonal elements $Q_{ij} (i \neq j)$ are mapped onto $J_{ij}$.
Constant terms do not affect the optimal solution and are thus omitted.
Several quantum algorithms can be used to determine the ground state of Eq.~\eqref{eq:ising}~\cite{Farhi2014,Finnila1994}.
Here, we focus on quantum annealing, which allows for slightly larger problem sizes than algorithms implemented on gate-based quantum computers.
The algorithm initializes the system in the ground state of a well-known and easily preparable Hamiltonian, then adjusts parameters to prepare the problem-encoding Hamiltonian Eq.~\eqref{eq:ising}.
If done slowly, the system will remain in the ground state.
However, quantum annealing implementations do not exclusively produce the lowest energy state but rather a probability distribution biased toward that state.
Sampling is then used to obtain near-optimal solutions for the \gls{qubo} problem.

The algorithm is implemented using the \textit{D-Wave's Advantage 4.1} \gls{qpu}, with over $5,\!000$ qubits, coupling each qubit with $15$ other~\cite{DWave2022}.
Here, each spin variable $\hat{\sigma}_j^z$ of the Ising Hamiltonian is mapped to a superconducting qubit on a lattice graph, and couplers implement interactions between qubits.
Mapping the full Ising Hamiltonian directly to the \gls{qpu} leads to long chains of redundant qubits, which are prone to noise.
Therefore, we utilize the hybrid quantum-classical algorithm \textit{QBSolv}~\cite{DWave2018}, combining partitioning of the \gls{qubo}-matrix and the heuristic Tabular-search algorithm.
The smaller sub-\glspl{qubo} are solved using the \textit{D-Wave} annealer.
This approach also allows solving larger instances of \gls{qubo} problems.

\subsection{Lagrange Parameter Search}
\label{subsec:langrange_parameter_search}

Introducing Lagrange parameters in Eq.~\eqref{eq:reformulation:lagrange} requires additional hyperparameter optimization.
These parameters balance the constraints' influence on the solution cost and their respective importance.
Their choice is highly problem-dependent and significantly affects solution quality.
If chosen too small, sampling could have a bias towards infeasible solutions; if chosen  too large, suboptimal solutions may predominate.
A grid search is performed to find the optimal Lagrange parameters.
For each combination, the \gls{qubo} is sampled $1,\!000$ times, selecting the combination with the highest number of optimal solutions.
A solution is considered optimal if it is valid, \ie, all constraints are satisfied, and the cost equals the global minimum of the \gls{ralbp}.
To limit \gls{qpu} execution time, we use simulated annealing from the \textit{dwave-neal} package~\cite{Neal2022} instead of direct quantum computer execution.
Simulated annealing is a classical algorithm, approximating global optima through probabilistic neighborhood search~\cite{Kirkpatrick1983}.
Using this algorithm here is reasonable because both quantum and classical annealing operate on the same cost landscape.
We assume that 1) an energy landscape that is favorable for simulated annealing benefits also quantum annealing, and 2) real quantum hardware errors are minor compared to Lagrange parameter modifications.

\section{Case Study}
\label{sec:case_study}

This section introduces a simple \gls{ralbp} case study on which the quantum annealing solution and the classical \gls{ip} model are compared.

\subsection{Data Description}
\label{subsec:data_description}

\begin{figure}[tbp]
    \centering
    \includegraphics[width=.4\columnwidth]{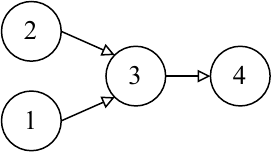}
    \caption{Precedence graph for the case study with $4$ tasks.}
    \label{fig:precedence_graph}
\end{figure}

We consider a reduced complexity \gls{ralbp} with four tasks across two workstations and two equipment types.
In the task matrix $t_{ij}$, $-1$ indicates inability to perform a specific task.
The task matrix $t_{ij}$ and cost vector $c_{j}$ are defined as
\begin{equation*}
	t_{ij} = \begin{pmatrix}
		8  & -1 \\
		13 & 14 \\
		18 & -1 \\
		15 & 20
	\end{pmatrix} \qquad
 	c_{j} = \begin{pmatrix}
			100,\!000 \\ 60,\!000
	\end{pmatrix} \, .
\end{equation*}
Finally, the directed precedence graph for the case study with $4$ tasks is shown in \autoref{fig:precedence_graph}, defining predecessor and successor relationships.
The cycle time constraint is set to $C=40$.

\subsection{Results}
\label{subsec:results}

The optimal assignment to the case study by an exact solution to the \gls{ip} model results in the task to equipment and workstation assignment shown in \autoref{tab:case_study_ip_solutions}.
The solution took $0.03\,\text{s}$ computation time.
\renewcommand{\arraystretch}{1.3}
\begin{table*}[htb]
	\centering
	\caption{The $C=40$ example took $0.03\,\text{s}$ computation time with an objective value of $160,\!000$, and $2$ pieces of equipment in $2$ workstations.}
	\label{tab:case_study_ip_solutions}
	\begin{tabularx}{.99\textwidth}{c*3{>{\centering\arraybackslash}X}cc}
		\hline
		Cycle time & Workstation & Equipment & Tasks              & Processing times                       & Workstation time \\
		\hline
		$C=40$     & $k=1$       & $j=1$     & $i = \{1, 2, 3 \}$ & $\{ t_{11}=8, t_{12}=13, t_{13}=18 \}$ & $39$             \\
		           & $k=2$       & $j=2$     & $i = \{ 4 \}$      & $\{ t_{24}=20 \}$                      & $20$             \\
		\hline
	\end{tabularx}
\end{table*}
\renewcommand{\arraystretch}{1.0}
The formulation as a \gls{qubo} yields a quadratic matrix of dimension $64 \times 64$, \ie, a problem with $64$ binary decision variables which can be encoded using $64$ qubits.
\begin{figure}
    \centering
    \includegraphics[width=\linewidth]{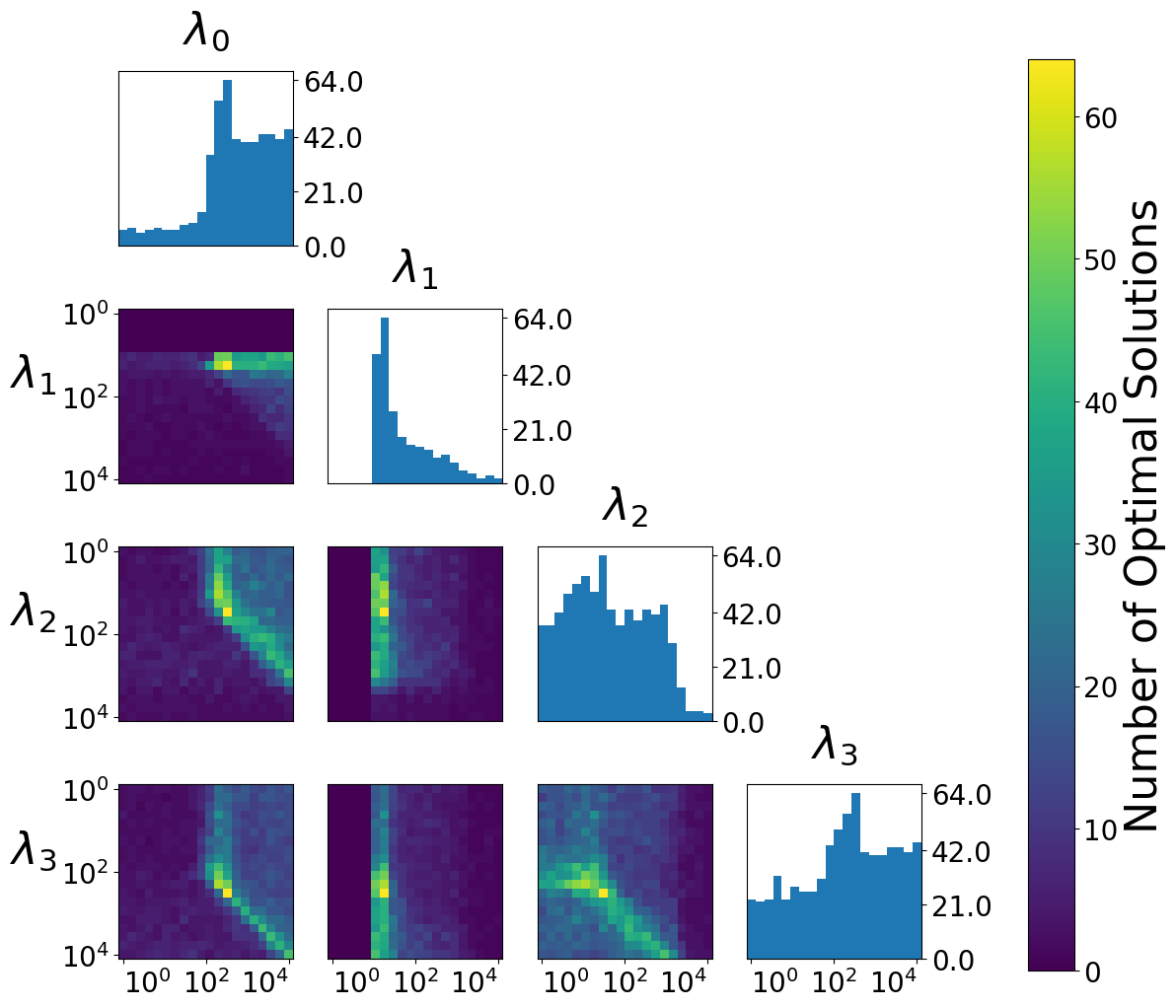}
    \caption{
		Number of optimal solutions for each fixed combination of Lagrange parameters in Eq.~\eqref{eq:reformulation:lagrange} out of $1,\!000$ samples with \textit{dwave-neal} simulated annealer.
		The plots on the diagonal show the maximal number of optimal solutions for a single fixed Lagrange parameter aggregated over all combinations of the other Lagrange parameters.
		The plots on the lower triangle show the maximal number of optimal solutions for two fixed Lagrange parameters.
		The set of optimal Lagrange parameters for our case study with \mbox{$C=40$} are $\lambda_1 = 335.982$, $\lambda_2 = 18.330$, $\lambda_3 = 29.764$, $\lambda_4 = 335.982$.
	}
    \label{fig:lpo-results}
\end{figure}
\autoref{fig:lpo-results} shows the search space and the results of the Lagrange parameter search from \autoref{subsec:langrange_parameter_search}.
Each row and column corresponds to a Lagrange parameter $\lambda_i$.
The histograms on the diagonal and the heatmaps on the lower triangle show the maximum number of occurrences when one or two Lagrange parameters are fixed.
While many configurations of Lagrange parameters can produce an optimal solution, there is a clear spike in the number of occurrences of the optimal solution visible in the histograms on the diagonal and the heatmaps on the lower triangle.

Using the optimal Lagrange parameters detailed in \autoref{fig:lpo-results}, we obtain $64$ optimal solutions out of $1,\!000$ samples with the \textit{dwave-neal} simulated annealer.
Sampling $1,\!000$ solutions with \textit{dwave-neal} took $4.3\,\text{s}$.
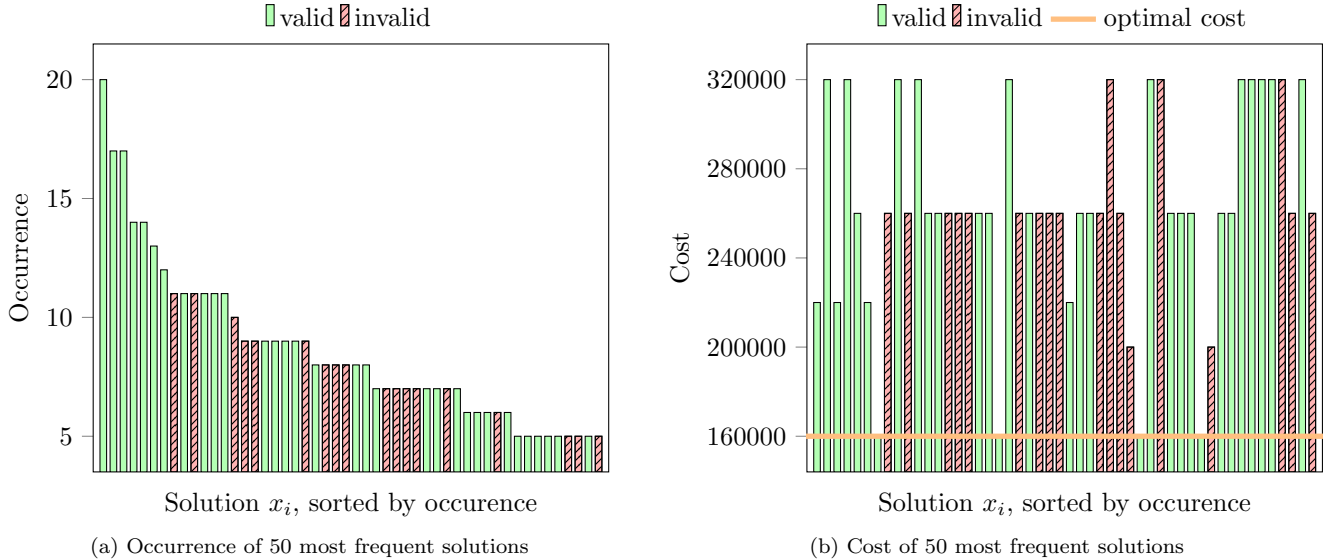
\begin{figure*}[htb]
    \centering
    \subfloat[\label{fig:dwave-results:a}Occurrence of 50 most frequent solutions]{
        \input{result_example_1_occurence.tex}
    }\hfill
    \subfloat[\label{fig:dwave-results:b}Cost of 50 most frequent solutions]{
        \input{result_example_1_cost.tex}
    }
    \caption{
		Top $50$ samples out of $1,\!000$ produced by \textit{QBSolv} with a \textit{D-Wave Advantage System 4.1} quantum annealer.
		Each bar represents one unique solution.
		The color and pattern of the bar are determined by whether the according solution is valid, \ie, fulfills all constraints Eqs.~\eqref{eq:constraint:task}-\eqref{eq:constraint:precedence}.
		In \protect\subref{fig:dwave-results:a}, the height of the bar shows the number of occurrences of the solution.
		In \protect\subref{fig:dwave-results:b}, the height of the bar shows the cost according to the original cost function in Eq.~\eqref{eq:objective}.
		The optimal cost value is also highlighted as a yellow line in \protect\subref{fig:dwave-results:b}.
    }
    \label{fig:dwave-results}
\end{figure*}
For the selected set of Lagrange parameters, we further run the \textit{QBSolv} algorithm on the \textit{D-Wave Advantage 4.1} \gls{qpu}, sampling $1,\!000$ solutions.
This sampling took a total computation time of $10,\!110.5\,\text{s}$, including the communication with the \gls{qpu} but excluding the computation of the Lagrange parameters, which are assumed to be known.

\autoref{fig:dwave-results:a} shows the $50$ most occurring solutions, ordered by occurrence and color-coded according to the validity of the solution.
\autoref{fig:dwave-results:b} shows the according cost value from Eq.~\eqref{eq:objective}.
Additionally, the optimal cost value, \ie, the cost value of the best possible solution, as obtained by the \gls{ip} solution, is marked in the plot.

Within the $50$ most occurring samples, there are $32$ valid solutions and $18$ invalid solutions.
Multiple optimal solutions are also occurring among the $50$ most occurring solutions.
The first optimal solution is in the seventh position and occurs $12$ times.
The $50$ most occurring samples contain $34$ occurrences of $4$ different configurations of an optimal solution.

Comparing the results from simulated and quantum annealing, we find that simulated annealing produces a larger number of optimal solutions than quantum annealing.
This is expected because quantum computers are subject to hardware noise and short coherence times, which ultimately require a trade-off between non-adiabatic errors if the annealing time is too fast and decoherence errors if it is too slow.

\section{Discussion}
\label{sec:discussion}

This work investigated the applicability of quantum annealing to the manufacturing optimization problem of \gls{ralb} to counteract the ever-increasing complexity in production domains, where traditional approaches struggle to solve problems efficiently.
A software toolbox was developed and published to transform the optimization instances into a \gls{qubo} formulation, making it solvable using quantum algorithms such as quantum annealing.
To avoid limitations of currently available quantum computers, a hybrid quantum-classical approach using \textit{QBSolv} in combination with a \textit{D-Wave Advantage 4.1} system \gls{qpu} was used to produce sample solutions.

We evaluated the application on a \gls{ralbp} case study instance, which was first transformed to a \gls{qubo} representation with the developed toolbox and then solved.
The results were compared to the classical \gls{ip} formulation and the simulated annealing algorithm.
All methods found the optimal solution, with the traditional \gls{ip} solution being the fastest.
Quantum annealing produced slightly less feasible solutions than simulated annealing.
However, simulated annealing is expected to get stuck more easily in local minima, which are more frequent with larger instances, ultimately hindering the scalability.
We point out that although we report and discuss execution times on our use case, a speed-up in computation time is neither the primary goal of this work nor to be expected in this problem instance.
The central premise is scaling advantages that might be harvested in larger instances.
However, the current state of quantum computing hardware limits the size of problems that can be investigated.
The low connectivity of the qubits within the quantum computer requires the use of ancillary qubits, which results in larger errors.
Ultimately, these errors currently limit the applicability to larger-scale \gls{ralbp} instances, which is why the current study is limited to this small use case.
With further progress of the quantum hardware, both problems could be mitigated by removing the need for the search-based hybrid algorithm and enabling the application to larger-scale \glspl{ralbp} where a scaling benefit is expected.
We thus present the results on a small instance as a step towards scalability, providing researchers with a starting point that includes the \gls{qubo}-transformation library, the showcased approach, and instance evaluation.
This foundation aims to enable future advancements in the field as quantum hardware continues to evolve and improve, ultimately leading to improved productivity and efficiency in manufacturing.

A significant benefit of using a sampling-based approach is the availability of multiple valid solutions rather than one, offsetting the computation time and, thus, scalability.
These solutions can be leveraged for metrics other than the original cost function without additional effort, \ie, no further sampling for more solutions is required.
A limitation of this approach is that the number of workstations has to be fixed for formulating the \gls{qubo}.
The choice regarding the predefined number of workstations must be conservative, providing enough workstations to produce a feasible solution, potentially requiring more qubits than are ultimately necessary.
Heuristics could provide a guided estimation for certain properties, such as the number of workstations.

Future investigations could include multi-objective optimization, leveraging the solution sampling for multiple production goals.
In general, the presented approach is promising for the manufacturing problem of \gls{ralb} if quantum annealing hardware advances to a point where it can be used efficiently for the full \gls{qubo} formulation.
Also, other quantum optimization approaches using gate-based quantum computers could lead to better results when more mature, especially fully fault-tolerant, quantum hardware is available.

\input{main.bbl}

\end{document}

%% file: acronyms.tex
\newacronym{alb}{ALB}{Assembly Line Balancing}
\newacronym{albp}{ALBP}{Assembly Line Balancing Problem}
\newacronym{cop}{COP}{combinatorial optimization problem}
\newacronym{cpu}{CPU}{Central Processing Unit}
\newacronym{ip}{IP}{Integer Programming}
\newacronym{qaoa}{QAOA}{Quantum Approximate Optimization Algorithm}
\newacronym{qpu}{QPU}{quantum processing unit}
\newacronym{qubo}{QUBO}{Quadratic Unconstrained Binary Optimization}
\newacronym{ralb}{RALB}{Robotic Assembly Line Balancing}
\newacronym{ralbp}{RALBP}{Robotic Assembly Line Balancing Problem}

\def\ie{\mbox{i.\,e.}}

%% file: result_example_1_occurence.tex
\begin{tikzpicture}
    \begin{axis}[
        ybar,
        bar width=2.5pt,
        bar shift=0mm,
        xmin=-1,
        xmax=50,
        xtick=\empty,
        xlabel={Solution $x_i$, sorted by occurence},
        ytick pos=left,
        ytick align=outside,
        ylabel={Occurrence},
        legend style={at={(0.5,1.11)},
            anchor=north,legend columns=-1,draw=none},
    ]
        \addplot[fill=green!30] coordinates {
            (0, 20)
            (1, 17)
            (2, 17)
            (3, 14)
            (4, 14)
            (5, 13)
            (6, 12)
            (8, 11)
            (10, 11)
            (11, 11)
            (12, 11)
            (16, 9)
            (17, 9)
            (18, 9)
            (19, 9)
            (21, 8)
            (25, 8)
            (26, 8)
            (27, 7)
            (32, 7)
            (33, 7)
            (35, 7)
            (36, 6)
            (37, 6)
            (38, 6)
            (40, 6)
            (41, 5)
            (42, 5)
            (43, 5)
            (44, 5)
            (45, 5)
            (48, 5)
        };
        \addplot[fill=red!30, postaction={pattern=north east lines}] coordinates {
            (7, 11)
            (9, 11)
            (13, 10)
            (14, 9)
            (15, 9)
            (20, 9)
            (22, 8)
            (23, 8)
            (24, 8)
            (28, 7)
            (29, 7)
            (30, 7)
            (31, 7)
            (34, 7)
            (39, 6)
            (46, 5)
            (47, 5)
            (49, 5)
        };
        \legend{valid, invalid}
    \end{axis}
\end{tikzpicture}

%% file: result_example_1_cost.tex
\begin{tikzpicture}
    \begin{axis}[
        ybar,
        bar width=2.5pt,
        bar shift=0mm,
        xmin=-1,
        xmax=50,
        xtick=\empty,
        xlabel={Solution $x_i$, sorted by occurence},
        ytick pos=left,
        ytick align=outside,
        yticklabel style={
            /pgf/number format/fixed,
            /pgf/number format/1000 sep={},
        },
        ytick={160000, 200000, 240000, 280000, 320000},
        scaled y ticks=false,
        ylabel={Cost},
        legend style={at={(0.5,1.11)},
            anchor=north,legend columns=-1,draw=none},
    ]
        \addplot[fill=green!30] coordinates {
            (0, 220000)
            (1, 320000)
            (2, 220000)
            (3, 320000)
            (4, 260000)
            (5, 220000)
            (6, 160000)
            (8, 320000)
            (10, 320000)
            (11, 260000)
            (12, 260000)
            (16, 260000)
            (17, 260000)
            (18, 160000)
            (19, 320000)
            (21, 260000)
            (25, 220000)
            (26, 260000)
            (27, 260000)
            (32, 160000)
            (33, 320000)
            (35, 260000)
            (36, 260000)
            (37, 260000)
            (38, 160000)
            (40, 260000)
            (41, 260000)
            (42, 320000)
            (43, 320000)
            (44, 320000)
            (45, 320000)
            (48, 320000)
        };
        \addplot[fill=red!30, postaction={pattern=north east lines}] coordinates {
            (7, 260000)
            (9, 260000)
            (13, 260000)
            (14, 260000)
            (15, 260000)
            (20, 260000)
            (22, 260000)
            (23, 260000)
            (24, 260000)
            (28, 260000)
            (29, 320000)
            (30, 260000)
            (31, 200000)
            (34, 320000)
            (39, 200000)
            (46, 320000)
            (47, 260000)
            (49, 260000)
        };
        \addplot[orange!50, line width=2pt, line legend, sharp plot, update limits=false] coordinates {
            (-10, 160000)
            (60, 160000)
        };
        \legend{valid, invalid, optimal cost}
    \end{axis}
\end{tikzpicture}

%% file: main.bbl
%

%% file: main.bbl
\begin{thebibliography}{31}%
\makeatletter
\providecommand \@ifxundefined [1]{%
 \@ifx{#1\undefined}
}%
\providecommand \@ifnum [1]{%
 \ifnum #1\expandafter \@firstoftwo
 \else \expandafter \@secondoftwo
 \fi
}%
\providecommand \@ifx [1]{%
 \ifx #1\expandafter \@firstoftwo
 \else \expandafter \@secondoftwo
 \fi
}%
\providecommand \natexlab [1]{#1}%
\providecommand \enquote  [1]{``#1''}%
\providecommand \bibnamefont  [1]{#1}%
\providecommand \bibfnamefont [1]{#1}%
\providecommand \citenamefont [1]{#1}%
\providecommand \href@noop [0]{\@secondoftwo}%
\providecommand \href [0]{\begingroup \@sanitize@url \@href}%
\providecommand \@href[1]{\@@startlink{#1}\@@href}%
\providecommand \@@href[1]{\endgroup#1\@@endlink}%
\providecommand \@sanitize@url [0]{\catcode `\\12\catcode `\$12\catcode
  `\&12\catcode `\#12\catcode `\^12\catcode `\_12\catcode `\%12\relax}%
\providecommand \@@startlink[1]{}%
\providecommand \@@endlink[0]{}%
\providecommand \url  [0]{\begingroup\@sanitize@url \@url }%
\providecommand \@url [1]{\endgroup\@href {#1}{\urlprefix }}%
\providecommand \urlprefix  [0]{URL }%
\providecommand \Eprint [0]{\href }%
\providecommand \doibase [0]{https://doi.org/}%
\providecommand \selectlanguage [0]{\@gobble}%
\providecommand \bibinfo  [0]{\@secondoftwo}%
\providecommand \bibfield  [0]{\@secondoftwo}%
\providecommand \translation [1]{[#1]}%
\providecommand \BibitemOpen [0]{}%
\providecommand \bibitemStop [0]{}%
\providecommand \bibitemNoStop [0]{.\EOS\space}%
\providecommand \EOS [0]{\spacefactor3000\relax}%
\providecommand \BibitemShut  [1]{\csname bibitem#1\endcsname}%
\let\auto@bib@innerbib\@empty
\bibitem [{\citenamefont {Becker}\ and\ \citenamefont
  {Scholl}(2006)}]{Becker2006}%
  \BibitemOpen
  \bibfield  {author} {\bibinfo {author} {\bibfnamefont {C.}~\bibnamefont
  {Becker}}\ and\ \bibinfo {author} {\bibfnamefont {A.}~\bibnamefont
  {Scholl}},\ }\bibfield  {title} {\bibinfo {title} {A survey on problems and
  methods in generalized assembly line balancing},\ }\href
  {https://doi.org/10.1016/j.ejor.2004.07.023} {\bibfield  {journal} {\bibinfo
  {journal} {European Journal of Operational Research}\ }\textbf {\bibinfo
  {volume} {168}},\ \bibinfo {pages} {694} (\bibinfo {year}
  {2006})}\BibitemShut {NoStop}%
\bibitem [{\citenamefont {Rubinovitz}\ \emph {et~al.}(1993)\citenamefont
  {Rubinovitz}, \citenamefont {Bukchin},\ and\ \citenamefont
  {Lenz}}]{Rubinovitz1993}%
  \BibitemOpen
  \bibfield  {author} {\bibinfo {author} {\bibfnamefont {J.}~\bibnamefont
  {Rubinovitz}}, \bibinfo {author} {\bibfnamefont {J.}~\bibnamefont
  {Bukchin}},\ and\ \bibinfo {author} {\bibfnamefont {E.}~\bibnamefont
  {Lenz}},\ }\bibfield  {title} {\bibinfo {title} {Ralb - a heuristic algorithm
  for design and balancing of robotic assembly lines},\ }\href
  {https://doi.org/10.1016/S0007-8506(07)62494-9} {\bibfield  {journal}
  {\bibinfo  {journal} {CIRP Annals - Manufacturing Technology}\ }\textbf
  {\bibinfo {volume} {42}},\ \bibinfo {pages} {497} (\bibinfo {year}
  {1993})}\BibitemShut {NoStop}%
\bibitem [{\citenamefont {Touzout}\ and\ \citenamefont
  {Benyoucef}(2019)}]{Touzout2019}%
  \BibitemOpen
  \bibfield  {author} {\bibinfo {author} {\bibfnamefont {F.~A.}\ \bibnamefont
  {Touzout}}\ and\ \bibinfo {author} {\bibfnamefont {L.}~\bibnamefont
  {Benyoucef}},\ }\bibfield  {title} {\bibinfo {title} {Multi-objective
  sustainable process plan generation in a reconfigurable manufacturing
  environment: Exact and adapted evolutionary approaches},\ }\href
  {https://doi.org/10.1080/00207543.2018.1522006} {\bibfield  {journal}
  {\bibinfo  {journal} {International Journal of Production Research}\ }\textbf
  {\bibinfo {volume} {57}},\ \bibinfo {pages} {2531} (\bibinfo {year}
  {2019})}\BibitemShut {NoStop}%
\bibitem [{\citenamefont {Riandari}\ \emph {et~al.}(2021)\citenamefont
  {Riandari}, \citenamefont {Alesha},\ and\ \citenamefont
  {Sihotang}}]{Riandari2021}%
  \BibitemOpen
  \bibfield  {author} {\bibinfo {author} {\bibfnamefont {F.}~\bibnamefont
  {Riandari}}, \bibinfo {author} {\bibfnamefont {A.}~\bibnamefont {Alesha}},\
  and\ \bibinfo {author} {\bibfnamefont {H.~T.}\ \bibnamefont {Sihotang}},\
  }\bibfield  {title} {\bibinfo {title} {Quantum computing for production
  planning},\ }\href {https://doi.org/10.35335/emod.v15i3.50} {\bibfield
  {journal} {\bibinfo  {journal} {International Journal of Enterprise
  Modelling}\ }\textbf {\bibinfo {volume} {15}},\ \bibinfo {pages} {163}
  (\bibinfo {year} {2021})}\BibitemShut {NoStop}%
\bibitem [{\citenamefont {Boysen}\ \emph {et~al.}(2022)\citenamefont {Boysen},
  \citenamefont {Schulze},\ and\ \citenamefont {Scholl}}]{Boysen2022}%
  \BibitemOpen
  \bibfield  {author} {\bibinfo {author} {\bibfnamefont {N.}~\bibnamefont
  {Boysen}}, \bibinfo {author} {\bibfnamefont {P.}~\bibnamefont {Schulze}},\
  and\ \bibinfo {author} {\bibfnamefont {A.}~\bibnamefont {Scholl}},\
  }\bibfield  {title} {\bibinfo {title} {{Assembly line balancing: What
  happened in the last fifteen years?}},\ }\href
  {https://doi.org/10.1016/J.EJOR.2021.11.043} {\bibfield  {journal} {\bibinfo
  {journal} {European Journal of Operational Research}\ }\textbf {\bibinfo
  {volume} {301}},\ \bibinfo {pages} {797} (\bibinfo {year}
  {2022})}\BibitemShut {NoStop}%
\bibitem [{\citenamefont {Luckow}\ \emph {et~al.}(2021)\citenamefont {Luckow},
  \citenamefont {Klepsch},\ and\ \citenamefont {Pichlmeier}}]{Luckow2021}%
  \BibitemOpen
  \bibfield  {author} {\bibinfo {author} {\bibfnamefont {A.}~\bibnamefont
  {Luckow}}, \bibinfo {author} {\bibfnamefont {J.}~\bibnamefont {Klepsch}},\
  and\ \bibinfo {author} {\bibfnamefont {J.}~\bibnamefont {Pichlmeier}},\
  }\bibfield  {title} {\bibinfo {title} {Quantum {{Computing}}: {{Towards
  Industry Reference Problems}}},\ }\href
  {https://doi.org/10.1007/s42354-021-0335-7} {\bibfield  {journal} {\bibinfo
  {journal} {Digitale Welt}\ }\textbf {\bibinfo {volume} {5}},\ \bibinfo
  {pages} {38} (\bibinfo {year} {2021})}\BibitemShut {NoStop}%
\bibitem [{\citenamefont {Pirnay}\ \emph {et~al.}(2024)\citenamefont {Pirnay},
  \citenamefont {Ulitzsch}, \citenamefont {Wilde}, \citenamefont {Eisert},\
  and\ \citenamefont {Seifert}}]{Pirnay2024}%
  \BibitemOpen
  \bibfield  {author} {\bibinfo {author} {\bibfnamefont {N.}~\bibnamefont
  {Pirnay}}, \bibinfo {author} {\bibfnamefont {V.}~\bibnamefont {Ulitzsch}},
  \bibinfo {author} {\bibfnamefont {F.}~\bibnamefont {Wilde}}, \bibinfo
  {author} {\bibfnamefont {J.}~\bibnamefont {Eisert}},\ and\ \bibinfo {author}
  {\bibfnamefont {J.-P.}\ \bibnamefont {Seifert}},\ }\bibfield  {title}
  {\bibinfo {title} {An in-principle super-polynomial quantum advantage for
  approximating combinatorial optimization problems via computational learning
  theory},\ }\href {https://doi.org/10.1126/sciadv.adj5170} {\bibfield
  {journal} {\bibinfo  {journal} {Science Advances}\ }\textbf {\bibinfo
  {volume} {10}},\ \bibinfo {pages} {eadj5170} (\bibinfo {year}
  {2024})}\BibitemShut {NoStop}%
\bibitem [{\citenamefont {Farhi}\ \emph {et~al.}(2014)\citenamefont {Farhi},
  \citenamefont {Goldstone},\ and\ \citenamefont {Gutmann}}]{Farhi2014}%
  \BibitemOpen
  \bibfield  {author} {\bibinfo {author} {\bibfnamefont {E.}~\bibnamefont
  {Farhi}}, \bibinfo {author} {\bibfnamefont {J.}~\bibnamefont {Goldstone}},\
  and\ \bibinfo {author} {\bibfnamefont {S.}~\bibnamefont {Gutmann}},\
  }\href@noop {} {\bibinfo {title} {A quantum approximate optimization
  algorithm}} (\bibinfo {year} {2014}),\ \Eprint
  {https://arxiv.org/abs/1411.4028} {arXiv:1411.4028 [quant-ph]} \BibitemShut
  {NoStop}%
\bibitem [{\citenamefont {Finnila}\ \emph {et~al.}(1994)\citenamefont
  {Finnila}, \citenamefont {Gomez}, \citenamefont {Sebenik}, \citenamefont
  {Stenson},\ and\ \citenamefont {Doll}}]{Finnila1994}%
  \BibitemOpen
  \bibfield  {author} {\bibinfo {author} {\bibfnamefont {A.}~\bibnamefont
  {Finnila}}, \bibinfo {author} {\bibfnamefont {M.}~\bibnamefont {Gomez}},
  \bibinfo {author} {\bibfnamefont {C.}~\bibnamefont {Sebenik}}, \bibinfo
  {author} {\bibfnamefont {C.}~\bibnamefont {Stenson}},\ and\ \bibinfo {author}
  {\bibfnamefont {J.}~\bibnamefont {Doll}},\ }\bibfield  {title} {\bibinfo
  {title} {Quantum annealing: A new method for minimizing multidimensional
  functions},\ }\href
  {https://doi.org/https://doi.org/10.1016/0009-2614(94)00117-0} {\bibfield
  {journal} {\bibinfo  {journal} {Chemical Physics Letters}\ }\textbf {\bibinfo
  {volume} {219}},\ \bibinfo {pages} {343} (\bibinfo {year}
  {1994})}\BibitemShut {NoStop}%
\bibitem [{\citenamefont {Abbas}\ \emph {et~al.}(2023)\citenamefont {Abbas},
  \citenamefont {Ambainis}, \citenamefont {Augustino}, \citenamefont
  {Bärtschi}, \citenamefont {Buhrman}, \citenamefont {Coffrin}, \citenamefont
  {Cortiana}, \citenamefont {Dunjko}, \citenamefont {Egger}, \citenamefont
  {Elmegreen}, \citenamefont {Franco}, \citenamefont {Fratini}, \citenamefont
  {Fuller}, \citenamefont {Gacon}, \citenamefont {Gonciulea}, \citenamefont
  {Gribling}, \citenamefont {Gupta}, \citenamefont {Hadfield}, \citenamefont
  {Heese}, \citenamefont {Kircher}, \citenamefont {Kleinert}, \citenamefont
  {Koch}, \citenamefont {Korpas}, \citenamefont {Lenk}, \citenamefont
  {Marecek}, \citenamefont {Markov}, \citenamefont {Mazzola}, \citenamefont
  {Mensa}, \citenamefont {Mohseni}, \citenamefont {Nannicini}, \citenamefont
  {O'Meara}, \citenamefont {Tapia}, \citenamefont {Pokutta}, \citenamefont
  {Proissl}, \citenamefont {Rebentrost}, \citenamefont {Sahin}, \citenamefont
  {Symons}, \citenamefont {Tornow}, \citenamefont {Valls}, \citenamefont
  {Woerner}, \citenamefont {Wolf-Bauwens}, \citenamefont {Yard}, \citenamefont
  {Yarkoni}, \citenamefont {Zechiel}, \citenamefont {Zhuk},\ and\ \citenamefont
  {Zoufal}}]{Abbas2023}%
  \BibitemOpen
  \bibfield  {author} {\bibinfo {author} {\bibfnamefont {A.}~\bibnamefont
  {Abbas}}, \bibinfo {author} {\bibfnamefont {A.}~\bibnamefont {Ambainis}},
  \bibinfo {author} {\bibfnamefont {B.}~\bibnamefont {Augustino}}, \bibinfo
  {author} {\bibfnamefont {A.}~\bibnamefont {Bärtschi}}, \bibinfo {author}
  {\bibfnamefont {H.}~\bibnamefont {Buhrman}}, \bibinfo {author} {\bibfnamefont
  {C.}~\bibnamefont {Coffrin}}, \bibinfo {author} {\bibfnamefont
  {G.}~\bibnamefont {Cortiana}}, \bibinfo {author} {\bibfnamefont
  {V.}~\bibnamefont {Dunjko}}, \bibinfo {author} {\bibfnamefont {D.~J.}\
  \bibnamefont {Egger}}, \bibinfo {author} {\bibfnamefont {B.~G.}\ \bibnamefont
  {Elmegreen}}, \bibinfo {author} {\bibfnamefont {N.}~\bibnamefont {Franco}},
  \bibinfo {author} {\bibfnamefont {F.}~\bibnamefont {Fratini}}, \bibinfo
  {author} {\bibfnamefont {B.}~\bibnamefont {Fuller}}, \bibinfo {author}
  {\bibfnamefont {J.}~\bibnamefont {Gacon}}, \bibinfo {author} {\bibfnamefont
  {C.}~\bibnamefont {Gonciulea}}, \bibinfo {author} {\bibfnamefont
  {S.}~\bibnamefont {Gribling}}, \bibinfo {author} {\bibfnamefont
  {S.}~\bibnamefont {Gupta}}, \bibinfo {author} {\bibfnamefont
  {S.}~\bibnamefont {Hadfield}}, \bibinfo {author} {\bibfnamefont
  {R.}~\bibnamefont {Heese}}, \bibinfo {author} {\bibfnamefont
  {G.}~\bibnamefont {Kircher}}, \bibinfo {author} {\bibfnamefont
  {T.}~\bibnamefont {Kleinert}}, \bibinfo {author} {\bibfnamefont
  {T.}~\bibnamefont {Koch}}, \bibinfo {author} {\bibfnamefont {G.}~\bibnamefont
  {Korpas}}, \bibinfo {author} {\bibfnamefont {S.}~\bibnamefont {Lenk}},
  \bibinfo {author} {\bibfnamefont {J.}~\bibnamefont {Marecek}}, \bibinfo
  {author} {\bibfnamefont {V.}~\bibnamefont {Markov}}, \bibinfo {author}
  {\bibfnamefont {G.}~\bibnamefont {Mazzola}}, \bibinfo {author} {\bibfnamefont
  {S.}~\bibnamefont {Mensa}}, \bibinfo {author} {\bibfnamefont
  {N.}~\bibnamefont {Mohseni}}, \bibinfo {author} {\bibfnamefont
  {G.}~\bibnamefont {Nannicini}}, \bibinfo {author} {\bibfnamefont
  {C.}~\bibnamefont {O'Meara}}, \bibinfo {author} {\bibfnamefont {E.~P.}\
  \bibnamefont {Tapia}}, \bibinfo {author} {\bibfnamefont {S.}~\bibnamefont
  {Pokutta}}, \bibinfo {author} {\bibfnamefont {M.}~\bibnamefont {Proissl}},
  \bibinfo {author} {\bibfnamefont {P.}~\bibnamefont {Rebentrost}}, \bibinfo
  {author} {\bibfnamefont {E.}~\bibnamefont {Sahin}}, \bibinfo {author}
  {\bibfnamefont {B.~C.~B.}\ \bibnamefont {Symons}}, \bibinfo {author}
  {\bibfnamefont {S.}~\bibnamefont {Tornow}}, \bibinfo {author} {\bibfnamefont
  {V.}~\bibnamefont {Valls}}, \bibinfo {author} {\bibfnamefont
  {S.}~\bibnamefont {Woerner}}, \bibinfo {author} {\bibfnamefont {M.~L.}\
  \bibnamefont {Wolf-Bauwens}}, \bibinfo {author} {\bibfnamefont
  {J.}~\bibnamefont {Yard}}, \bibinfo {author} {\bibfnamefont {S.}~\bibnamefont
  {Yarkoni}}, \bibinfo {author} {\bibfnamefont {D.}~\bibnamefont {Zechiel}},
  \bibinfo {author} {\bibfnamefont {S.}~\bibnamefont {Zhuk}},\ and\ \bibinfo
  {author} {\bibfnamefont {C.}~\bibnamefont {Zoufal}},\ }\href@noop {}
  {\bibinfo {title} {Quantum optimization: Potential, challenges, and the path
  forward}} (\bibinfo {year} {2023}),\ \Eprint
  {https://arxiv.org/abs/2312.02279} {arXiv:2312.02279 [quant-ph]} \BibitemShut
  {NoStop}%
\bibitem [{\citenamefont {King}\ \emph {et~al.}(2023)\citenamefont {King},
  \citenamefont {Raymond}, \citenamefont {Lanting}, \citenamefont {Harris},
  \citenamefont {Zucca}, \citenamefont {Altomare}, \citenamefont {Berkley},
  \citenamefont {Boothby}, \citenamefont {Ejtemaee}, \citenamefont {Enderud},
  \citenamefont {Hoskinson}, \citenamefont {Huang}, \citenamefont {Ladizinsky},
  \citenamefont {MacDonald}, \citenamefont {Marsden}, \citenamefont {Molavi},
  \citenamefont {Oh}, \citenamefont {{Poulin-Lamarre}}, \citenamefont {Reis},
  \citenamefont {Rich}, \citenamefont {Sato}, \citenamefont {Tsai},
  \citenamefont {Volkmann}, \citenamefont {Whittaker}, \citenamefont {Yao},
  \citenamefont {Sandvik},\ and\ \citenamefont {Amin}}]{King2023}%
  \BibitemOpen
  \bibfield  {author} {\bibinfo {author} {\bibfnamefont {A.~D.}\ \bibnamefont
  {King}}, \bibinfo {author} {\bibfnamefont {J.}~\bibnamefont {Raymond}},
  \bibinfo {author} {\bibfnamefont {T.}~\bibnamefont {Lanting}}, \bibinfo
  {author} {\bibfnamefont {R.}~\bibnamefont {Harris}}, \bibinfo {author}
  {\bibfnamefont {A.}~\bibnamefont {Zucca}}, \bibinfo {author} {\bibfnamefont
  {F.}~\bibnamefont {Altomare}}, \bibinfo {author} {\bibfnamefont {A.~J.}\
  \bibnamefont {Berkley}}, \bibinfo {author} {\bibfnamefont {K.}~\bibnamefont
  {Boothby}}, \bibinfo {author} {\bibfnamefont {S.}~\bibnamefont {Ejtemaee}},
  \bibinfo {author} {\bibfnamefont {C.}~\bibnamefont {Enderud}}, \bibinfo
  {author} {\bibfnamefont {E.}~\bibnamefont {Hoskinson}}, \bibinfo {author}
  {\bibfnamefont {S.}~\bibnamefont {Huang}}, \bibinfo {author} {\bibfnamefont
  {E.}~\bibnamefont {Ladizinsky}}, \bibinfo {author} {\bibfnamefont {A.~J.~R.}\
  \bibnamefont {MacDonald}}, \bibinfo {author} {\bibfnamefont {G.}~\bibnamefont
  {Marsden}}, \bibinfo {author} {\bibfnamefont {R.}~\bibnamefont {Molavi}},
  \bibinfo {author} {\bibfnamefont {T.}~\bibnamefont {Oh}}, \bibinfo {author}
  {\bibfnamefont {G.}~\bibnamefont {{Poulin-Lamarre}}}, \bibinfo {author}
  {\bibfnamefont {M.}~\bibnamefont {Reis}}, \bibinfo {author} {\bibfnamefont
  {C.}~\bibnamefont {Rich}}, \bibinfo {author} {\bibfnamefont {Y.}~\bibnamefont
  {Sato}}, \bibinfo {author} {\bibfnamefont {N.}~\bibnamefont {Tsai}}, \bibinfo
  {author} {\bibfnamefont {M.}~\bibnamefont {Volkmann}}, \bibinfo {author}
  {\bibfnamefont {J.~D.}\ \bibnamefont {Whittaker}}, \bibinfo {author}
  {\bibfnamefont {J.}~\bibnamefont {Yao}}, \bibinfo {author} {\bibfnamefont
  {A.~W.}\ \bibnamefont {Sandvik}},\ and\ \bibinfo {author} {\bibfnamefont
  {M.~H.}\ \bibnamefont {Amin}},\ }\bibfield  {title} {\bibinfo {title}
  {Quantum critical dynamics in a 5,000-qubit programmable spin glass},\ }\href
  {https://doi.org/10.1038/s41586-023-05867-2} {\bibfield  {journal} {\bibinfo
  {journal} {Nature}\ }\textbf {\bibinfo {volume} {617}},\ \bibinfo {pages}
  {61} (\bibinfo {year} {2023})}\BibitemShut {NoStop}%
\bibitem [{\citenamefont {Willmann}(2024)}]{alb_qubo_code}%
  \BibitemOpen
  \bibfield  {author} {\bibinfo {author} {\bibfnamefont {M.}~\bibnamefont
  {Willmann}},\ }\href
  {https://gitlab.cc-asp.fraunhofer.de/ipa-quantum/alb-qubo} {\bibinfo {title}
  {{ALB QUBO} (version 1.0.0)}} (\bibinfo {year} {2024})\BibitemShut {NoStop}%
\bibitem [{\citenamefont {Li}\ \emph {et~al.}(2020)\citenamefont {Li},
  \citenamefont {Kucukkoc},\ and\ \citenamefont {Tang}}]{Li2020}%
  \BibitemOpen
  \bibfield  {author} {\bibinfo {author} {\bibfnamefont {Z.}~\bibnamefont
  {Li}}, \bibinfo {author} {\bibfnamefont {I.}~\bibnamefont {Kucukkoc}},\ and\
  \bibinfo {author} {\bibfnamefont {Q.}~\bibnamefont {Tang}},\ }\bibfield
  {title} {\bibinfo {title} {A comparative study of exact methods for the
  simple assembly line balancing problem},\ }\href
  {https://doi.org/10.1007/s00500-019-04609-9} {\bibfield  {journal} {\bibinfo
  {journal} {Soft Computing}\ }\textbf {\bibinfo {volume} {24}},\ \bibinfo
  {pages} {11459} (\bibinfo {year} {2020})}\BibitemShut {NoStop}%
\bibitem [{\citenamefont {Zhang}\ \emph {et~al.}(2018)\citenamefont {Zhang},
  \citenamefont {Hu},\ and\ \citenamefont {Wu}}]{Zhang2018}%
  \BibitemOpen
  \bibfield  {author} {\bibinfo {author} {\bibfnamefont {Y.}~\bibnamefont
  {Zhang}}, \bibinfo {author} {\bibfnamefont {X.}~\bibnamefont {Hu}},\ and\
  \bibinfo {author} {\bibfnamefont {C.}~\bibnamefont {Wu}},\ }\bibfield
  {title} {\bibinfo {title} {Heuristic algorithm for type ii two-sided assembly
  line rebalancing problem with multi-objective},\ }in\ \href
  {https://doi.org/10.1051/matecconf/201817503063} {\emph {\bibinfo {booktitle}
  {MATEC Web of Conferences}}},\ Vol.\ \bibinfo {volume} {175}\ (\bibinfo
  {year} {2018})\ p.\ \bibinfo {pages} {03063}\BibitemShut {NoStop}%
\bibitem [{\citenamefont {Borba}\ \emph {et~al.}(2018)\citenamefont {Borba},
  \citenamefont {Ritt},\ and\ \citenamefont {Miralles}}]{Borba2018}%
  \BibitemOpen
  \bibfield  {author} {\bibinfo {author} {\bibfnamefont {L.}~\bibnamefont
  {Borba}}, \bibinfo {author} {\bibfnamefont {M.}~\bibnamefont {Ritt}},\ and\
  \bibinfo {author} {\bibfnamefont {C.}~\bibnamefont {Miralles}},\ }\bibfield
  {title} {\bibinfo {title} {Exact and heuristic methods for solving the
  robotic assembly line balancing problem},\ }\href
  {https://doi.org/10.1016/j.ejor.2018.03.011} {\bibfield  {journal} {\bibinfo
  {journal} {European Journal of Operational Research}\ }\textbf {\bibinfo
  {volume} {270}},\ \bibinfo {pages} {146} (\bibinfo {year}
  {2018})}\BibitemShut {NoStop}%
\bibitem [{\citenamefont {Albus}\ \emph {et~al.}(2024)\citenamefont {Albus},
  \citenamefont {Hornek}, \citenamefont {Kraus},\ and\ \citenamefont
  {Huber}}]{Albus2024}%
  \BibitemOpen
  \bibfield  {author} {\bibinfo {author} {\bibfnamefont {M.}~\bibnamefont
  {Albus}}, \bibinfo {author} {\bibfnamefont {T.}~\bibnamefont {Hornek}},
  \bibinfo {author} {\bibfnamefont {W.}~\bibnamefont {Kraus}},\ and\ \bibinfo
  {author} {\bibfnamefont {M.~F.}\ \bibnamefont {Huber}},\ }\bibfield  {title}
  {\bibinfo {title} {Towards scalability for resource reconfiguration in
  robotic assembly line balancing problems using a modified genetic
  algorithm},\ }\bibfield  {journal} {\bibinfo  {journal} {Journal of
  Intelligent Manufacturing}\ }\href
  {https://doi.org/10.1007/s10845-023-02292-0} {10.1007/s10845-023-02292-0}
  (\bibinfo {year} {2024})\BibitemShut {NoStop}%
\bibitem [{\citenamefont {Manavizadeh}\ \emph {et~al.}(2012)\citenamefont
  {Manavizadeh}, \citenamefont {Rabbani}, \citenamefont {Moshtaghi},\ and\
  \citenamefont {Jolai}}]{Manavizadeh2012}%
  \BibitemOpen
  \bibfield  {author} {\bibinfo {author} {\bibfnamefont {N.}~\bibnamefont
  {Manavizadeh}}, \bibinfo {author} {\bibfnamefont {M.}~\bibnamefont
  {Rabbani}}, \bibinfo {author} {\bibfnamefont {D.}~\bibnamefont {Moshtaghi}},\
  and\ \bibinfo {author} {\bibfnamefont {F.}~\bibnamefont {Jolai}},\ }\bibfield
   {title} {\bibinfo {title} {Mixed-model assembly line balancing in the
  make-to-order and stochastic environment using multi-objective evolutionary
  algorithms},\ }\href {https://doi.org/10.1016/j.eswa.2012.03.044} {\bibfield
  {journal} {\bibinfo  {journal} {Expert Systems with Applications}\ }\textbf
  {\bibinfo {volume} {39}},\ \bibinfo {pages} {12026} (\bibinfo {year}
  {2012})}\BibitemShut {NoStop}%
\bibitem [{\citenamefont {Chutima}(2022)}]{Chutima2022}%
  \BibitemOpen
  \bibfield  {author} {\bibinfo {author} {\bibfnamefont {P.}~\bibnamefont
  {Chutima}},\ }\bibfield  {title} {\bibinfo {title} {A comprehensive review of
  robotic assembly line balancing problem},\ }\href
  {https://doi.org/10.1007/s10845-020-01641-7} {\bibfield  {journal} {\bibinfo
  {journal} {Journal of Intelligent Manufacturing}\ }\textbf {\bibinfo {volume}
  {33}},\ \bibinfo {pages} {1} (\bibinfo {year} {2022})}\BibitemShut {NoStop}%
\bibitem [{\citenamefont {Klar}\ \emph {et~al.}(2022)\citenamefont {Klar},
  \citenamefont {Schworm}, \citenamefont {Wu}, \citenamefont {Glatt},\ and\
  \citenamefont {Aurich}}]{Klar2022}%
  \BibitemOpen
  \bibfield  {author} {\bibinfo {author} {\bibfnamefont {M.}~\bibnamefont
  {Klar}}, \bibinfo {author} {\bibfnamefont {P.}~\bibnamefont {Schworm}},
  \bibinfo {author} {\bibfnamefont {X.}~\bibnamefont {Wu}}, \bibinfo {author}
  {\bibfnamefont {M.}~\bibnamefont {Glatt}},\ and\ \bibinfo {author}
  {\bibfnamefont {J.~C.}\ \bibnamefont {Aurich}},\ }\bibfield  {title}
  {\bibinfo {title} {Quantum {{Annealing}} based factory layout planning},\
  }\href {https://doi.org/10.1016/j.mfglet.2022.03.003} {\bibfield  {journal}
  {\bibinfo  {journal} {Manufacturing Letters}\ }\textbf {\bibinfo {volume}
  {32}},\ \bibinfo {pages} {59} (\bibinfo {year} {2022})}\BibitemShut {NoStop}%
\bibitem [{\citenamefont {Dolgui}\ and\ \citenamefont
  {Ihnatsenka}(2009)}]{Dolgui2009}%
  \BibitemOpen
  \bibfield  {author} {\bibinfo {author} {\bibfnamefont {A.}~\bibnamefont
  {Dolgui}}\ and\ \bibinfo {author} {\bibfnamefont {I.}~\bibnamefont
  {Ihnatsenka}},\ }\bibfield  {title} {\bibinfo {title} {Branch and bound
  algorithm for a transfer line design problem: Stations with sequentially
  activated multi-spindle heads},\ }\href
  {https://doi.org/10.1016/j.ejor.2008.03.028} {\bibfield  {journal} {\bibinfo
  {journal} {European Journal of Operational Research}\ }\textbf {\bibinfo
  {volume} {197}},\ \bibinfo {pages} {1119} (\bibinfo {year}
  {2009})}\BibitemShut {NoStop}%
\bibitem [{\citenamefont {Yarkoni}\ \emph {et~al.}(2022)\citenamefont
  {Yarkoni}, \citenamefont {Raponi}, \citenamefont {B{\"a}ck},\ and\
  \citenamefont {Schmitt}}]{Yarkoni2022}%
  \BibitemOpen
  \bibfield  {author} {\bibinfo {author} {\bibfnamefont {S.}~\bibnamefont
  {Yarkoni}}, \bibinfo {author} {\bibfnamefont {E.}~\bibnamefont {Raponi}},
  \bibinfo {author} {\bibfnamefont {T.}~\bibnamefont {B{\"a}ck}},\ and\
  \bibinfo {author} {\bibfnamefont {S.}~\bibnamefont {Schmitt}},\ }\bibfield
  {title} {\bibinfo {title} {Quantum annealing for industry applications:
  Introduction and review},\ }\href {https://doi.org/10.1088/1361-6633/ac8c54}
  {\bibfield  {journal} {\bibinfo  {journal} {Reports on Progress in Physics}\
  }\textbf {\bibinfo {volume} {85}},\ \bibinfo {pages} {104001} (\bibinfo
  {year} {2022})}\BibitemShut {NoStop}%
\bibitem [{\citenamefont {Cohen}\ and\ \citenamefont
  {Tamir}(2014)}]{Cohen2014}%
  \BibitemOpen
  \bibfield  {author} {\bibinfo {author} {\bibfnamefont {E.}~\bibnamefont
  {Cohen}}\ and\ \bibinfo {author} {\bibfnamefont {B.}~\bibnamefont {Tamir}},\
  }\bibfield  {title} {\bibinfo {title} {D-{{Wave}} and predecessors: {{From}}
  simulated to quantum annealing},\ }\href
  {https://doi.org/10.1142/S0219749914300022} {\bibfield  {journal} {\bibinfo
  {journal} {International Journal of Quantum Information}\ }\textbf {\bibinfo
  {volume} {12}},\ \bibinfo {pages} {1430002} (\bibinfo {year}
  {2014})}\BibitemShut {NoStop}%
\bibitem [{\citenamefont {Liu}\ and\ \citenamefont {Li}(2021)}]{Liu2021}%
  \BibitemOpen
  \bibfield  {author} {\bibinfo {author} {\bibfnamefont {Z.}~\bibnamefont
  {Liu}}\ and\ \bibinfo {author} {\bibfnamefont {S.}~\bibnamefont {Li}},\
  }\bibfield  {title} {\bibinfo {title} {A {{Quantum Computing Based Numerical
  Method}} for {{Solving Mixed-Integer Optimal Control Problems}}},\ }\href
  {https://doi.org/10.1007/s11424-020-9278-6} {\bibfield  {journal} {\bibinfo
  {journal} {Journal of Systems Science and Complexity}\ }\textbf {\bibinfo
  {volume} {34}},\ \bibinfo {pages} {2428} (\bibinfo {year}
  {2021})}\BibitemShut {NoStop}%
\bibitem [{\citenamefont {Kurowski}\ \emph {et~al.}(2020)\citenamefont
  {Kurowski}, \citenamefont {W{\c e}glarz}, \citenamefont {Subocz},
  \citenamefont {R{\'o}{\.z}ycki},\ and\ \citenamefont
  {Walig{\'o}ra}}]{Kurowski2020}%
  \BibitemOpen
  \bibfield  {author} {\bibinfo {author} {\bibfnamefont {K.}~\bibnamefont
  {Kurowski}}, \bibinfo {author} {\bibfnamefont {J.}~\bibnamefont {W{\c
  e}glarz}}, \bibinfo {author} {\bibfnamefont {M.}~\bibnamefont {Subocz}},
  \bibinfo {author} {\bibfnamefont {R.}~\bibnamefont {R{\'o}{\.z}ycki}},\ and\
  \bibinfo {author} {\bibfnamefont {G.}~\bibnamefont {Walig{\'o}ra}},\
  }\bibfield  {title} {\bibinfo {title} {Hybrid {{Quantum Annealing Heuristic
  Method}} for {{Solving Job Shop Scheduling Problem}}},\ }in\ \href
  {https://doi.org/10.1007/978-3-030-50433-5_39} {\emph {\bibinfo {booktitle}
  {Computational {{Science}} -- {{ICCS}} 2020}}},\ Vol.\ \bibinfo {volume}
  {12142}\ (\bibinfo  {publisher} {Springer International Publishing},\
  \bibinfo {address} {Cham},\ \bibinfo {year} {2020})\ pp.\ \bibinfo {pages}
  {502--515}\BibitemShut {NoStop}%
\bibitem [{\citenamefont {Schworm}\ \emph {et~al.}(2023)\citenamefont
  {Schworm}, \citenamefont {Wu}, \citenamefont {Klar}, \citenamefont {Gayer},
  \citenamefont {Glatt},\ and\ \citenamefont {Aurich}}]{Schworm2023}%
  \BibitemOpen
  \bibfield  {author} {\bibinfo {author} {\bibfnamefont {P.}~\bibnamefont
  {Schworm}}, \bibinfo {author} {\bibfnamefont {X.}~\bibnamefont {Wu}},
  \bibinfo {author} {\bibfnamefont {M.}~\bibnamefont {Klar}}, \bibinfo {author}
  {\bibfnamefont {J.}~\bibnamefont {Gayer}}, \bibinfo {author} {\bibfnamefont
  {M.}~\bibnamefont {Glatt}},\ and\ \bibinfo {author} {\bibfnamefont {J.~C.}\
  \bibnamefont {Aurich}},\ }\bibfield  {title} {\bibinfo {title} {Resilience
  optimization in manufacturing systems using {{Quantum Annealing}}},\ }\href
  {https://doi.org/10.1016/j.mfglet.2022.12.007} {\bibfield  {journal}
  {\bibinfo  {journal} {Manufacturing Letters}\ }\textbf {\bibinfo {volume}
  {36}},\ \bibinfo {pages} {13} (\bibinfo {year} {2023})}\BibitemShut {NoStop}%
\bibitem [{\citenamefont {Glos}\ \emph {et~al.}(2023)\citenamefont {Glos},
  \citenamefont {Kundu},\ and\ \citenamefont {Salehi}}]{Glos2023}%
  \BibitemOpen
  \bibfield  {author} {\bibinfo {author} {\bibfnamefont {A.}~\bibnamefont
  {Glos}}, \bibinfo {author} {\bibfnamefont {A.}~\bibnamefont {Kundu}},\ and\
  \bibinfo {author} {\bibfnamefont {{\"O}.}~\bibnamefont {Salehi}},\ }\bibfield
   {title} {\bibinfo {title} {Optimizing the {{Production}} of {{Test Vehicles
  Using Hybrid Constrained Quantum Annealing}}},\ }\href
  {https://doi.org/10.1007/s42979-023-02071-x} {\bibfield  {journal} {\bibinfo
  {journal} {SN Computer Science}\ }\textbf {\bibinfo {volume} {4}},\ \bibinfo
  {pages} {609} (\bibinfo {year} {2023})}\BibitemShut {NoStop}%
\bibitem [{\citenamefont {Lucas}(2014)}]{Lucas2014}%
  \BibitemOpen
  \bibfield  {author} {\bibinfo {author} {\bibfnamefont {A.}~\bibnamefont
  {Lucas}},\ }\bibfield  {title} {\bibinfo {title} {Ising formulations of many
  np problems},\ }\bibfield  {journal} {\bibinfo  {journal} {Frontiers in
  Physics}\ }\textbf {\bibinfo {volume} {2}},\ \href
  {https://doi.org/10.3389/fphy.2014.00005} {10.3389/fphy.2014.00005} (\bibinfo
  {year} {2014})\BibitemShut {NoStop}%
\bibitem [{\citenamefont {McGeoch}\ and\ \citenamefont
  {Farr\'{e}}(2022)}]{DWave2022}%
  \BibitemOpen
  \bibfield  {author} {\bibinfo {author} {\bibfnamefont {C.}~\bibnamefont
  {McGeoch}}\ and\ \bibinfo {author} {\bibfnamefont {P.}~\bibnamefont
  {Farr\'{e}}},\ }\href
  {https://www.dwavesys.com/media/s3qbjp3s/14-1049a-a_the_d-wave_advantage_system_an_overview.pdf}
  {\emph {\bibinfo {title} {Advantage Processor Overview}}},\ \bibinfo {type}
  {Tech. Rep.}\ \bibinfo {number} {14-1058A-A}\ (\bibinfo  {institution}
  {D-Wave Systems Inc.},\ \bibinfo {year} {2022})\BibitemShut {NoStop}%
\bibitem [{\citenamefont {Booth}\ \emph {et~al.}(2017)\citenamefont {Booth},
  \citenamefont {Reinhardt},\ and\ \citenamefont {Roy}}]{DWave2018}%
  \BibitemOpen
  \bibfield  {author} {\bibinfo {author} {\bibfnamefont {M.}~\bibnamefont
  {Booth}}, \bibinfo {author} {\bibfnamefont {S.~P.}\ \bibnamefont
  {Reinhardt}},\ and\ \bibinfo {author} {\bibfnamefont {A.}~\bibnamefont
  {Roy}},\ }\href
  {https://www.dwavesys.com/media/jhlpvult/partitioning_qubos_for_quantum_acceleration-2.pdf}
  {\emph {\bibinfo {title} {Partitioning Optimization Problems for Hybrid
  Classical/Quantum Execution}}},\ \bibinfo {type} {Tech. Rep.}\ \bibinfo
  {number} {14-1006A-A}\ (\bibinfo  {institution} {D-Wave Systems Inc.},\
  \bibinfo {year} {2017})\BibitemShut {NoStop}%
\bibitem [{\citenamefont {{D-Wave Systens Inc.}}(2022)}]{Neal2022}%
  \BibitemOpen
  \bibfield  {author} {\bibinfo {author} {\bibnamefont {{D-Wave Systens
  Inc.}}},\ }\href {https://github.com/dwavesystems/dwave-neal} {\bibinfo
  {title} {dwave-neal (version 0.6.0)}} (\bibinfo {year} {2022})\BibitemShut
  {NoStop}%
\bibitem [{\citenamefont {Kirkpatrick}\ \emph {et~al.}(1983)\citenamefont
  {Kirkpatrick}, \citenamefont {Gelatt},\ and\ \citenamefont
  {Vecchi}}]{Kirkpatrick1983}%
  \BibitemOpen
  \bibfield  {author} {\bibinfo {author} {\bibfnamefont {S.}~\bibnamefont
  {Kirkpatrick}}, \bibinfo {author} {\bibfnamefont {C.~D.}\ \bibnamefont
  {Gelatt}},\ and\ \bibinfo {author} {\bibfnamefont {M.~P.}\ \bibnamefont
  {Vecchi}},\ }\bibfield  {title} {\bibinfo {title} {Optimization by simulated
  annealing},\ }\href {https://doi.org/10.1126/science.220.4598.671} {\bibfield
   {journal} {\bibinfo  {journal} {Science}\ }\textbf {\bibinfo {volume}
  {220}},\ \bibinfo {pages} {671} (\bibinfo {year} {1983})}\BibitemShut
  {NoStop}%
\end{thebibliography}
